\documentclass[prb,twocolumn,showpacs,preprintnumbers,amsmath,amssymb,floats,longbibliography]{revtex4-1}
\pdfoutput=1
\usepackage{graphicx}
\usepackage{amsmath}
\usepackage{epstopdf}
\usepackage{amsbsy}
\usepackage{color}
\usepackage{hyperref}
\usepackage{dcolumn}
\usepackage{bm}

\newcommand\R{{\mathbf R}}
\newcommand{\BSCCO}{{Bi$_2$Sr$_2$CaCu$_2$O$_8$ }}
\newcommand{\NaCCOC}{{Ca$_{2-x}$Na$_x$CuO$_2$Cl$_2$}}
\newcommand{\CCOC}{{Ca$_{2}$CuO$_2$Cl$_2$}}

\begin{document}
\title{Universality of scanning tunneling microscopy  in cuprate superconductors}
\author{Peayush Choubey$^{1,2}$, Andreas Kreisel$^{3,4}$, T. Berlijn$^{5}$, Brian M. Andersen$^4$ and  P. J. Hirschfeld$^1$}
\affiliation{
$^1$Department of Physics, University of Florida, Gainesville, Florida 32611, USA\\
$^2$Department of Physics, Indian Institute of Science, Bengaluru 560012, India\\
$^3$Institut f\" ur Theoretische Physik, Universit\" at Leipzig, D-04103 Leipzig, Germany\\
$^4$Niels Bohr Institute, Juliane Maries Vej 30, University of Copenhagen, DK-2100 Copenhagen, Denmark\\
$^5$Center for Nanophase Materials Sciences and Computational Sciences and Engineering Division, Oak Ridge National Laboratory, Oak Ridge, Tennessee 37831, USA
}

\date{\today}

\begin{abstract}
We consider the problem of local tunneling into cuprate superconductors, combining model based calculations for the superconducting order parameter with
wavefunction information obtained from first principles electronic structure.  For some time it has been proposed  that scanning tunneling microscopy (STM) spectra do not reflect the properties of the superconducting layer in the CuO$_2$ plane directly beneath the STM tip, but rather a weighted sum of spatially proximate states determined by the details of the tunneling process.  These ``filter'' ideas have been countered with the argument that similar conductance patterns have been seen around impurities and charge ordered states in systems with atomically quite different barrier layers.
Here we use a recently developed Wannier function based method to calculate topographies, spectra, conductance maps and normalized conductance maps close to impurities.
We find that it is  the local planar Cu $d_{x^2-y^2}$ Wannier function, qualitatively similar for many systems, that controls the form of the tunneling spectrum and the spatial patterns near perturbations.
We explain how, despite the fact that 
STM observables  depend on the materials-specific details of the tunneling process and setup parameters, there is an overall universality in the qualitative features of conductance spectra.  In particular, we discuss why  STM results on  \BSCCO (BSCCO) and \NaCCOC (NaCCOC) are essentially identical.  
\end{abstract}

\pacs{74.20.-z, 74.70.Xa, 74.62.En, 74.81.-g}

\maketitle
\section{Introduction}
\begin{figure}
\includegraphics[width=\columnwidth]{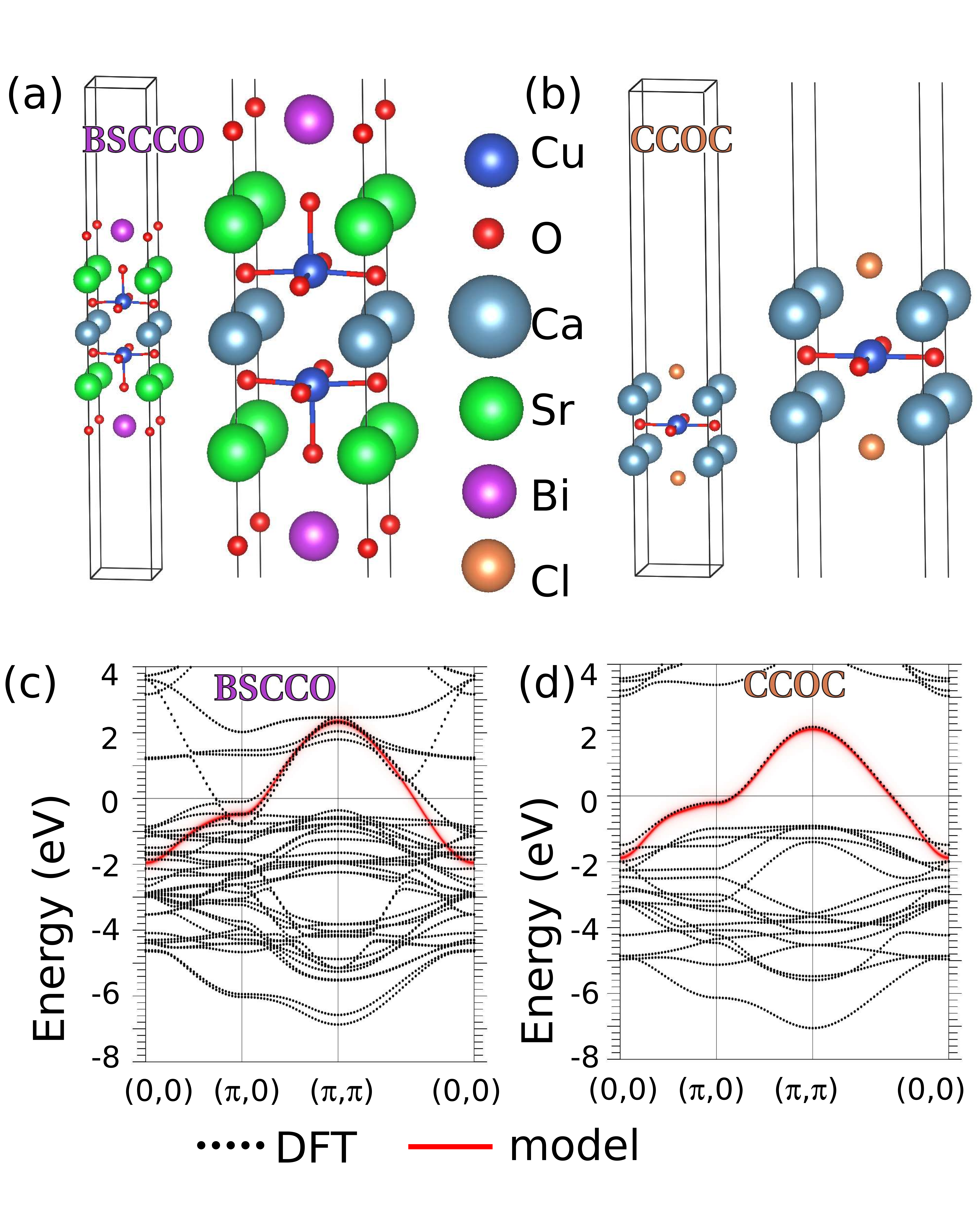}
\caption{Crystal structure of (a) Bi$_2$Sr$_2$CaCu$_2$O$_8$ and (b) Ca$_2$CuO$_2$Cl$_2$ surface as used for the DFT calculation. Band structure (black dots) of (c) Bi$_2$Sr$_2$CaCu$_2$O$_8$ and (d) Ca$_2$CuO$_2$Cl$_2$ surface compared to downfolding to a one-band model (red solid line).}
\label{fig_crystal_BSCCO}
\end{figure}

Scanning tunneling microscopy (STM) and spectroscopy (STS) have played an important role in the evolution of ideas about cuprate superconductors, including the $d$-wave symmetry of the superconductivity, the nature of the pseudogap, and the existence of competing orders\cite{FischerRMP,FujitaJPSJ}.   On the other hand, exactly what is measured by STM/S has never been completely clear, due to the fact that the superconducting wavefunction is believed to reside primarily in the CuO$_2$ layer, while the STM tip detects {a tunneling current ascribed to} the density of electronic states several \AA ~above the cleaved surface.  Between the two lie insulating barrier layers that are different for each cuprate material.  For example, in the canonical STM material \BSCCO (BSCCO), the insulating layer consists of BiO and SrO planes, with the latter containing a so-called ``apical'' O atom placed directly above a Cu in the CuO$_2$ plane below the  BiO layer at which the system cleaves.    In another intensively studied material, \NaCCOC (NaCCOC), the barrier layer contains planes of Cl and planes of Ca/Na atoms, with no apical O, see Fig. \ref{fig_crystal_BSCCO} (a)-(b). Instead, NaCCOC contains apical Cl, which reside on the surface of the cleaved layer.  If the tunneling process depends on the details of the barrier layer, one might expect to see very different results for different cuprate materials, even if their superconducting states are quite similar.  In practice, most theoretical work has ignored this complication and assumed a featureless barrier. 

To reveal the effects of the barrier layer most clearly, a homogeneous situation is not ideal, since the range of the tunneling process is hidden by the crystal periodicity\cite{Markiewicz}. On the other hand, studies of impurity effects in superconductors have been shown to be very useful in elucidating aspects of the superconductivity \cite{BalatskyRMP,AlloulRMP}.
In cuprates, Zn and Ni impurities have been widely studied as they substitute an in-plane Cu atom and hence directly couple to the electronic states that give rise to the superconducting properties. With a closed d-shell, Zn acts as a strong non-magnetic potential scatterer, whereas, Ni acts as a magnetic scatterer owing to its $3d^{8}$ configuration. The STM tunneling conductance around Zn in BSCCO features a sharp in-gap virtual bound state and drastic suppression of the coherence peaks\cite{PanZn}, consistent with the simplest theoretical predictions for a  nonmagnetic $\delta$-function potential \cite{Balatsky95} in a $d$-wave superconductor. However, the spatial pattern around the impurity site deviates from the predictions of this simple model\cite{BalatskyRMP}. Most notably, experiment observes intensity maxima at the impurity sites in contrast to the minima predicted by the theory.

To  reconcile  theory and experiment, 
several authors focused on the details of the tunneling process, and advanced a hypothesis that we will refer to as a ``filter''; in essence, these works proposed phenomenologically that the tip detected the signal arising from the four nearest neighbor Cu sites, rather than from that immediately below the tip\cite{Martin,Ting}.  The authors of Ref. \onlinecite{Martin} suggested, in particular, that the overlap of the apical oxygen 2$p_z$ and 3s orbitals with Bi $p_z$ was crucial for the tunneling.   While a first principles calculation of a Zn impurity in BSCCO detected a weak filter effect\cite{abinitioZn}, the effect of superconductivity was unknown.  More recently, a hybrid theory was proposed\cite{Choubey2014,Kreisel15} that accounted for material-specific details by  downfolding density functional theory calculations onto a tight-binding model using Cu 3$d_{x^2-y^2}$ Wannier functions, and combined this  with a phenomenological Bogoliubov-de Gennes (BdG) calculation of  $d$-wave superconductivity in this basis.  This ``BdG+W'' theory was able to account for the details of the Zn resonance in BSCCO and many aspects of the quasiparticle interference data on this material\cite{Kreisel15}. Furthermore, using a variant of the same method applicable to $t-J$ type models, it was recently shown \cite{Choubey2017} that the characteristics of $d$-form factor charge order, where charge modulations occur mainly at O sites and are opposite in phase for two inequivalent O atoms \cite{Fujita14,Comin2016,Hamidian2015}, can be easily obtained within a one-band calculation by accounting for the O degrees of freedom through the Wannier function. 

Nevertheless, while this model successfully accounted for   STM data on Zn impurities in BSCCO as well as quasiparticle interference results in exquisite detail,  it confirmed that the wavefunctions at the surface detected by the tip were influenced strongly by their hybridization with the apical oxygen 2$p_z$ states.  As discussed above, these apical O states are present in most cuprates, but not in NaCCOC. The previous Wannier-based work thus fails to address the question of the apparent universality of STM observables, e.g. impurity states and charge order, in different cuprate materials\cite{Davis_checkerboard2}. 
In this paper, we attempt to reconcile the filter proposals with the universality of STM response observed in cuprates.

For this purpose, we apply the BdG+W approach to the  impurity problem in both  BSCCO and NaCCOC.  We review the BSCCO Zn results, and move on to discuss the Ni impurity results from BSCCO, which were not addressed in Ref. \onlinecite{Kreisel15}.  STM studies on BSCCO \cite{Hudson2001} have shown two spin-resolved in-gap virtual bound states at energies $\pm\Omega_{1,2}^{\text{expt}}$. Bound state peaks were observed to be particle-like at the impurity site and next nearest neighbor sites, and hole like at the nearest neighbor sites. The spatial patterns at $+\Omega_{1,2}^{\text{expt}}$ resembled cross-shaped and X shaped at $-\Omega_{1,2}^{\text{expt}}$. Observation of these resonance states is consistent with the models of combined potential and magnetic scattering of quasiparticles in $d$-wave superconductors\cite{Salkola}. However, as in the case of Zn, spatial patterns deviate from the predictions\cite{Salkola}. In one attempt to reconcile theory and experiment ``filter'' was proposed\cite{Martin}. Here, we show that the experimental features are easily obtained in the BdG+W approach and, like the Zn problem\cite{Kreisel15}, tunneling via nearest neighbor apical oxygen atoms plays a crucial role in shaping the spatial patterns at resonance energies in the local density of states (LDOS) spectrum near Ni impurities.

For the same problem in NaCCOC, we find that the Wannier functions on the surface that couple to the STM tip and the electronic structure at low energies are very similar to BSCCO despite the fact that the material has a different structure and the tunneling takes place through a rather different surface layer. Consequently, the spectral properties and the spatial pattern observed close to strong potential scatterers and also magnetic scatterers are expected to be very similar to those seen in BSCCO.  We discuss the implications for observation of charge order in both materials, and for the use of STM to deduce important aspects of superconductivity in the cuprates.

\section{Model}
The method used in this work was introduced earlier\cite{Choubey2014} and already applied to  the case of cuprate superconductors already\cite{Kreisel15} as well as in multiband systems\cite{Choubey2014,Chi16}. The starting point is a first principles calculation that yields the electronic structure downfolded to a tight-binding model
\begin{equation}
 H_{\text{TB}}= Z \sum_{\mathbf{R},\mathbf{R'}\sigma} t^{\mu\nu}_{\mathbf{R},\mathbf{R'}} c_{\mathbf R \mu  \sigma}^\dagger c_{\mathbf R' \nu \sigma}-\mu_0\sum_{\mathbf{R \sigma}}c_{\mathbf R \mu  \sigma}^\dagger c_{\mathbf R \mu \sigma},
 \label{eq_tb}
\end{equation}
where $t_{\mathbf{R},\mathbf{R'}}^{\mu\nu}$ are hopping elements between orbitals $\mu$ and $\nu$ on the lattice sites labeled with $\mathbf{R}$ and $\mathbf{R'}$
and $c^\dagger_{\R\sigma}$ creates an electron at lattice site $\R$  with spin $\sigma = \pm$.
To account for correlations, we introduce an overall band renormalization $Z$ and work away from half filling using a rigid band shift by choosing the chemical potential $\mu_0$ accordingly. While this is a standard procedure and also a common starting point for model based calculations, we obtain extra information with the Wannier functions that describe the electrons. The field operators of the electrons are related to the lattice operators via
\begin{equation}
\psi_\sigma(\mathbf{r})=\sum_{\mathbf{R} \mu } c_{\mathbf{R}\mu\sigma}w_{\mathbf{R}\mu}(\mathbf{r}),
\label{eq_cont_trafo}
\end{equation}
where the Wannier functions $w_{\mathbf{R}\mu}(\mathbf{r})$ are the matrix elements.   For application of this approach to cuprates, we restrict consideration to a single Cu orbital, $\mu=\nu=d_{x^2-y^2}$.

In the following we study STM imaging of magnetic impurity states like Ni in cuprates. A magnetic impurity acts as the source of potential as well as magnetic scattering, and can be simply modeled by the following Hamiltonian
\begin{equation}
\begin{aligned}
H_{\text{imp}} = &\sum_{\R\sigma}\left(U_{\R^{\star}\R}c^\dagger_{\R^{\star}\sigma}c_{\R\sigma} + {\mathrm{H.c.}}\right) \\
&+\sum_{\R\sigma}\left(\sigma J_{\R^{\star}\R}c^\dagger_{\R^{\star}\sigma}c_{\R\sigma} + {\mathrm{ H.c.}}\right).
\end{aligned}
\label{Eq:Himp}
\end{equation}
The first term in the above Hamiltonian accounts for the potential scattering at the impurity site $\R^{\star}$ and $U_{\R^{\star}\R}$ is the impurity potential which can be written as $U_{0}\delta_{\R^{\star}\R}$, where $\delta$ represents the Kronecker delta function, for a completely local impurity model. The second term in the above Hamiltonian accounts for the magnetic scattering due to the exchange interaction between the impurity magnetic moment, approximated as a classical spin, and the conduction electrons. $J_{\R^{\star}\R}$ denotes the extended exchange coupling which can be expressed as $J_{0}\delta_{\R^{\star}\R}$ for the special case of completely local exchange interaction. Finally, to account for superconductivity we include a mean-field BCS term
\begin{equation}
 H_{\text{BCS}} = \sum_{\R,\R'}\left(\Delta_{\R,\R'}c^\dagger_{\R\uparrow}c^\dagger_{\R'\downarrow} + {\mathrm {H.c.}}\right),
\end{equation}
where $\Delta_{\R,\R'}$ are the pairing mean fields.

\section{Theoretical approach}

\subsection{T-matrix approach}
\label{sec_tmatrix}
For this purpose, we Fourier transform the superconducting order parameter of the homogeneous system to obtain $\Delta(\mathbf{k})$. In this work, we assume  a standard $d$-wave order parameter $\Delta_{\mathbf k}=\Delta_0(\cos(k_x)-\cos(k_y))$, and write down the Nambu Hamiltonian
\begin{equation}
 { H}_{\text{N}}(\mathbf{k})=\left(\begin{array}{cc}
               H(\mathbf{k})& \Delta(\mathbf{k})\\
               \Delta(\mathbf{k})^\dagger&- H(\mathbf{-k})^T
                  \end{array}\right),
                  \label{eq_H_nambu}
\end{equation}
where the Fourier transform $H(\mathbf{k})$  of the hopping elements has been introduced. Defining a Green's function of the non-interacting electronic structure via
\begin{equation}
{G}_{\mathbf{k}}^0(\omega)=[H_{\text{N}}(\mathbf{k})-\omega+i0^+]^{-1},
\label{eq_GF}
\end{equation}
we transform to real space
\begin{equation}
 { G}_{\mathbf{R}, \mathbf{R'}}^0(\omega)=\sum_{\mathbf{k}}e^{-i\mathbf{k}\cdot(\mathbf{R}-\mathbf{R'})}
{ G}_{\mathbf{k}}^0(\omega)={ G}_{\mathbf{R}-\mathbf{R'}}^0(\omega)\,,\label{eq_lattice_GF}
\end{equation}
in order to calculate the Green's function in the presence of the impurity
\begin{equation}
 { G}_{\mathbf{R}, \mathbf{R'}}(\omega)={ G}_{\mathbf{R}-\mathbf{R'}}^0(\omega)+{ G}_{\mathbf{R}}^0(\omega){ T}(\omega){ G}_{-\mathbf{R'}}^0(\omega)\,,
 \label{eq_lattice_GF_T}
\end{equation}
using the T-matrix
\begin{equation}
 { T}(\omega)=[1-{ V}_{\text{N,imp}}{ G}(\omega)]^{-1} { V}_{\text{N,imp}}\,.
 \label{eq_T_matrix}
\end{equation}
Here ${G}(\omega)=\sum_{\mathbf{k}}{G}_{\mathbf{k}}^0(\omega)$ is the local Green's function and the impurity potential, split up in a potential part and a magnetic contribution, is given by
$ V_{\text{N,imp}}=\sigma_z \otimes U_0+1_2 \otimes J_0$.
The steps presented up to now, are straightforward and widely used in the literature \cite{BalatskyRMP}; alternative methods to obtain the lattice Green's function ${ G}_{\mathbf{R}, \mathbf{R'}}(\omega)$ including the modulation of the order parameter in real space (via self-consistency approaches) will be presented in the next section. For the results presented in Sec. \ref{Sec:Results}, we employ the T-matrix approach only for the computationally expensive calculations of STM topographs and related maps.   
\subsection{BdG approach}
The mean-field solution of our problem including all terms of the Hamiltonian
\begin{equation}
 H= H_{\text{TB}}+ H_{\text{BCS}}+H_{\text{imp}},
 \label{Eq:Hamiltonain}
\end{equation}
can be found by solving a self-consistency equation where the superconducting order parameter $\Delta_{\R,\R'}$ is related to the pair potential $\Gamma_{\R,\R'}$ through $\Delta_{\R,\R'} = \Gamma_{\R,\R'}\left\langle c_{\R'\downarrow}c_{\R\uparrow}\right\rangle$. To get the $d$-wave gap symmetry we set $\Gamma_{\R,\R'} = \Gamma$ if sites $(\R,\R')$ are nearest neighbors and $0$ otherwise. The eigenvalue problem corresponding to Eq. (\ref{Eq:Hamiltonain}) can be expressed in the form of BdG equations which is solved self-consistently\cite{Choubey2014} for $\Delta_{\R,\R'}$ and $\mu_{0}$ by keeping the electron number fixed. Using BdG eigenvalues $E_{n\sigma}$ and eigenvectors $u_{n\sigma}$ and $v_{n\sigma}$, the lattice Green's function $G_{\R\R'\sigma}(\omega)$ can be obtained as well.

\subsection{Calculation of the tunneling conductance}
The differential tunneling conductance as measured in an STM experiment at a given bias voltage $V$  can be calculated using\cite{Tersoff1985}
\begin{equation}
\frac{dI}{dV}(\mathbf{r},eV)=\frac{4\pi e}{\hbar} \rho_t(0) |M|^2 \rho(\mathbf{r},eV),
\label{eq_conductance}
\end{equation}
where the continuum LDOS is given by
\begin{equation}
 \rho(\mathbf{r},\omega)\equiv -\frac 2 \pi\mathop\text{Im}{G}^{11}(\mathbf{r},\mathbf{r};\omega ),
 \label{eq_cLDOS}
\end{equation}
where ${G}^{11}$ refers to the particle-hole component of the Nambu Green's function; for broken $SU(2)$ symmetry in spin space, one needs to take into account also the ${G}^{22}$ component at negative energies
\begin{equation}
\rho(\mathbf{r},\omega)\equiv -\frac 1 \pi[\mathop\text{Im}{G}^{11}(\mathbf{r},\mathbf{r};\omega )-\mathop\text{Im}{G}^{22}(\mathbf{r},\mathbf{r};-\omega ).
 \label{eq_cLDOSm}
\end{equation}
Finally, the continuum Green's function can be calculated by employing a Wannier basis transformation\cite{Choubey2014},
\begin{equation}
 {G}(\mathbf{r},\mathbf{r}';\omega)=\sum_{\mathbf{R}, \mathbf{R'},\mu\nu}{ G}_{\mathbf{R}, \mathbf{R'}}^{\mu,\nu}(\omega)w_{\mathbf{R}\mu}(\mathbf{r})w_{\mathbf{R'}\nu}^{\ast}(\mathbf{r'}),
 \label{eq_basis}
\end{equation}
where relation Eq. (\ref{eq_cont_trafo}) has been used.
Note that the framework can be applied to single and multiband systems, while in the following, all orbital indices are ignored since a one-band model for cuprates is used.

\section{Results}
\label{Sec:Results}
\subsection{Electronic structure}

\begin{figure}
\includegraphics[width=1\columnwidth]{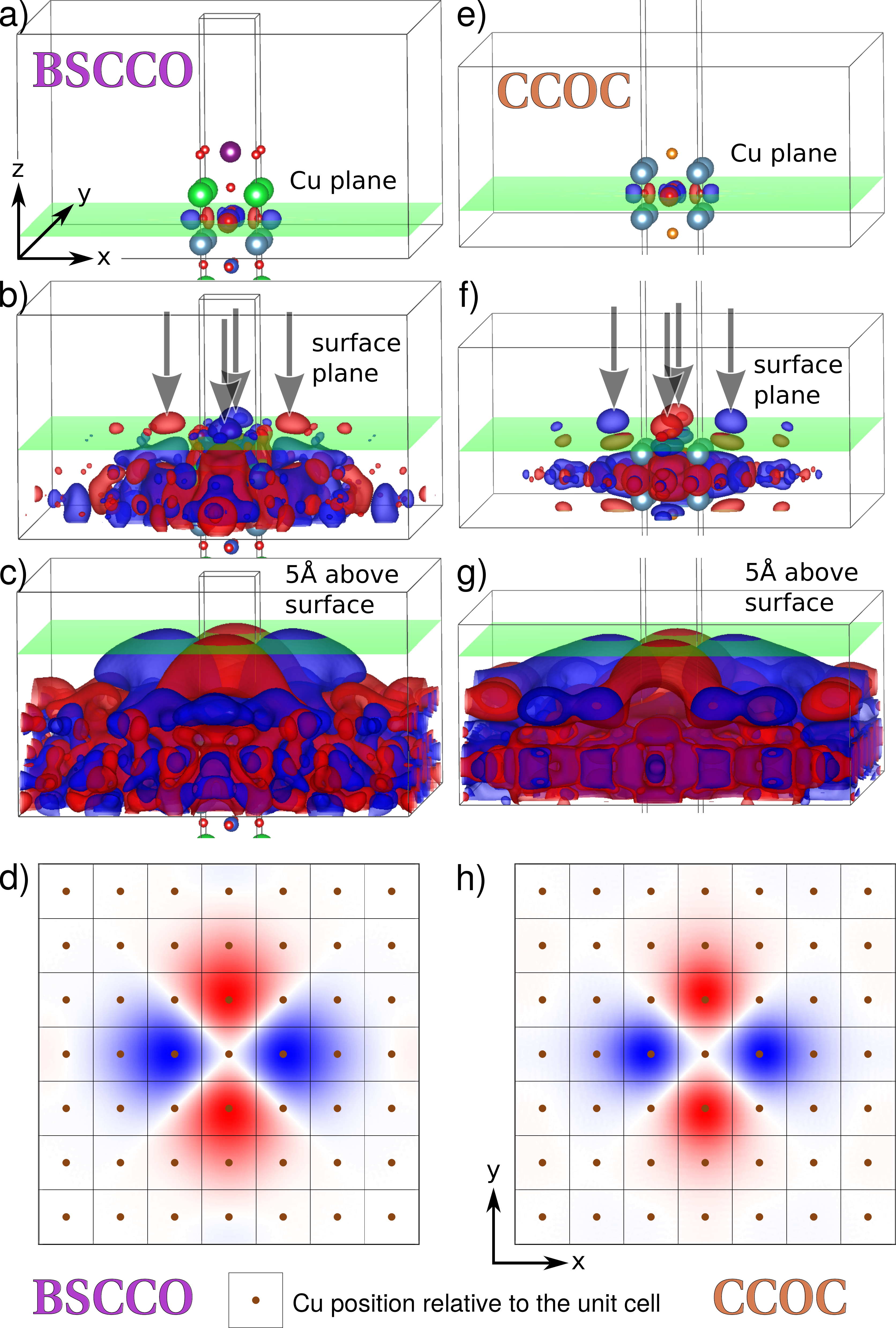}
\caption{Wannier function of (a-d) Bi$_2$Sr$_2$CaCu$_2$O$_8$ and (e-h) Ca$_2$CuO$_2$Cl$_2$ plotted at three different isovalues and as a colormap approximately 5 {\AA} above the surface exposed to the STM tip, as used for the simulations of STM conductance maps and spectra. The green planes in the isosurface plots lie in the plane of the Cu atoms, the plane of the surface atoms and 5 {\AA} above the surface atoms respectively.  The lobes of the Wannier function originating from hybridization with orbitals of the surface atoms are indicated by arrows in (b, f).  These correspond to apical O and apical Cl $p_z$ states in (b) and (f), respectively. The isovalues in $\text{bohr}^{-3/2}$ are the following $5\cdot 10^{-2}$ (a,e), $5\cdot 10^{-4}$ (b), $5\cdot 10^{-3}$ (f), $5\cdot 10^{-5}$ (c), $3\cdot 10^{-5}$ (g). The coordinate system used throughout this manuscript is indicated as inset in (a).}
\label{fig_wannier}
\end{figure}

The first step in the Wannier function based analysis of STM observables is to construct the tight-binding description of the normal state of the material under study, using first principles calculations.  As input for the density functional theory (DFT) calculations, a  cell representing the  Bi$_2$Sr$_2$CaCu$_2$O$_8$ surface was constructed as follows. First the bulk Bi$_4$Sr$_4$Ca$_2$Cu$_4$O$_{16}$ unit cell was considered with the structural parameters from Ref. \onlinecite{Hybertsen1988}. Then half of the atoms were deleted, resulting in a single Bi$_2$Sr$_2$CaCu$_2$O$_8$ layer unit cell with a vacuum of approximately 20 \AA, see Fig. \ref{fig_crystal_BSCCO} (a). A single Ca$_2$CuO$_2$Cl$_2$ (CCOC) layer unit cell was created in a similar way. First the bulk  Ca$_4$CuO$_4$Cl$_4$ unit cell was considered with structural parameters taken from Ref. \onlinecite{Argyriou95}. Then again half of the atoms were deleted which creates a certain amount of vacuum. In addition, some vacuum was added such that the total vacuum is again approximately 20 \AA. The DFT calculations and subsequent downfolding to the Wannier basis, were performed using the VASP\cite{Kresse} and Wannier90 \cite{Mostofi2014} packages, respectively.   After the construction of the tight-binding model, calculations are performed for 15\% doping implemented by a rigid band shift, roughly appropriate for optimal doping.

As shown in Fig. \ref{fig_crystal_BSCCO} (c)-(d), we obtain a very similar tight binding structure for both BSCCO and CCOC, which is generic for many cuprate materials.  The Wannier functions generated in this downfolding procedure involve  states derived from all atomic species in the unit cell, but restricted to an energy range of a few eV around the Fermi level.  The Cu $d_{x^2-y^2}$ Wannier function important for superconductivity is shown in Fig. \ref{fig_wannier} to involve not only atomic $d_{x^2-y^2}$ functions on a given site, but states localized on many nearby atoms.   The tunneling process depends on the density of these states, and those associated with Wannier functions on neighboring Cu's, at a point several \AA ~ above the surface.  At intermediate isovalues, i.e. distances from a given Cu, the Wannier function is seen in Fig. \ref{fig_wannier} to be rather complex, and differences between  BSCCO and CCOC are evident, reflecting difference in the barrier layer.    While in both cases, apical $p_z$ states above the nearest neighbor Cu are clearly visible, in the BSCCO case these are associated with O, while in  CCOC they are associated with the Cl.  There are also qualitative differences that may be observed upon close inspection of the figure.   

However, in the asymptotic regime far from the surface,  the  Wannier functions for both materials are seen to be almost identical. The symmetry of the Cu $d_{x^2-y^2}$ wavefunction is the same as dictated by the symmetry of the crystal including the atoms in the layers above the Cu-O layer leading to the vanishing weight directly above the central Cu site; small differences in the extent of the wavefunction, e.g. the positions of the maxima relative to the center, can be observed, see Fig. \ref{fig_wannier} (d),(h). The maximum in the magnitude of the Wannier function in both compounds occur above the NN Cu sites. It was shown in Ref. \onlinecite{Kreisel15} that this maximum has the largest contribution from the apical O-$p$ states in NN unit cells. In case of CCOC, we find that the NN Cl-$p$ states at the surface are responsible for the Wannier function maxima. The similarity of the shape of the surface Wannier function in both compounds leads to the similar spatial structure of impurity states at the STM tip position, as we show in following sections. 

\subsection{Effects of a non-magnetic impurity}

\begin{figure}
\includegraphics[width=1\columnwidth]{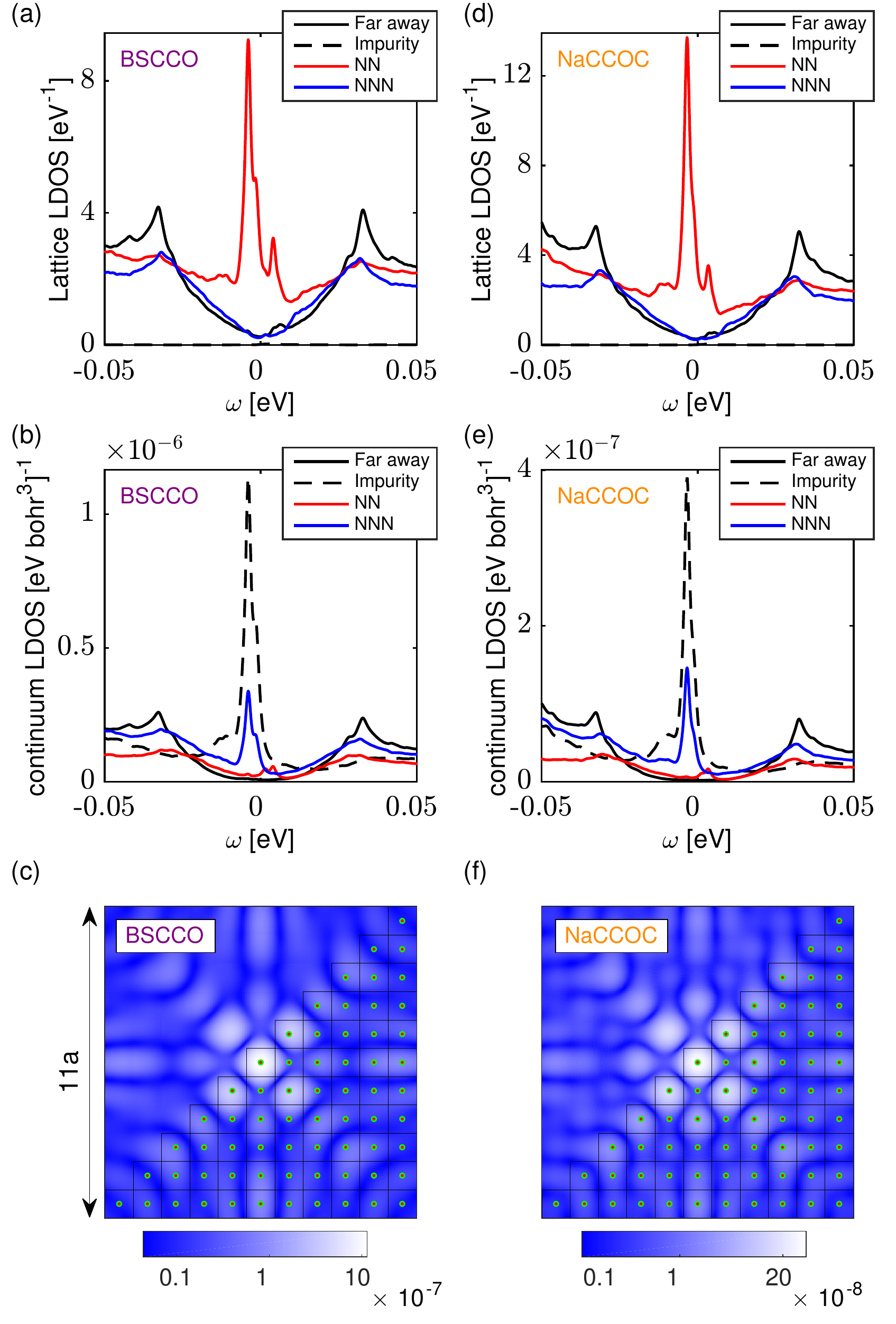}
\caption{
(a) Lattice LDOS spectrum around a strong, non-magnetic, Zn-like impurity, replacing Cu in BSCCO, with on-site potential $U_{0} = -5$ eV. Spectrum at a site far from the impurity (black), impurity site (dashed), nearest neighbor site (red), and next-nearest neighbor site (blue) is calculated using $30\times30$ supercells with artificial broadening of 1 meV. (b) Continuum LDOS spectrum at a height approximately 5 {\AA} above the BiO surface. Shown are positions directly above a Cu atom far from impurity (black), at the impurity (dashed), on the nearest neighbor position (red), and on the next-nearest neighbor position (blue). (c) Continuum LDOS map in the vicinity of impurity obtained at the resonance energy $\Omega = -4$ meV and a height approximately 5 {\AA} above the BiO surface. Colorbar values are in the units of eV$^{-1}$bohr$^{-3}$. Small black squares denote CuO$_2$ unit cells, with Cu positions marked by red-green circles. (d),(e),(f) same quantities as in (a),(b),(c), respectively, for the impurity replacing Cu in NaCCOC. The LDOS map in (e) is obtained at the resonance energy $\Omega = -3.4$ meV.}
\label{ldos_strong_impurity}
\end{figure}

\begin{figure}[tb]
\includegraphics[width=0.99\linewidth]{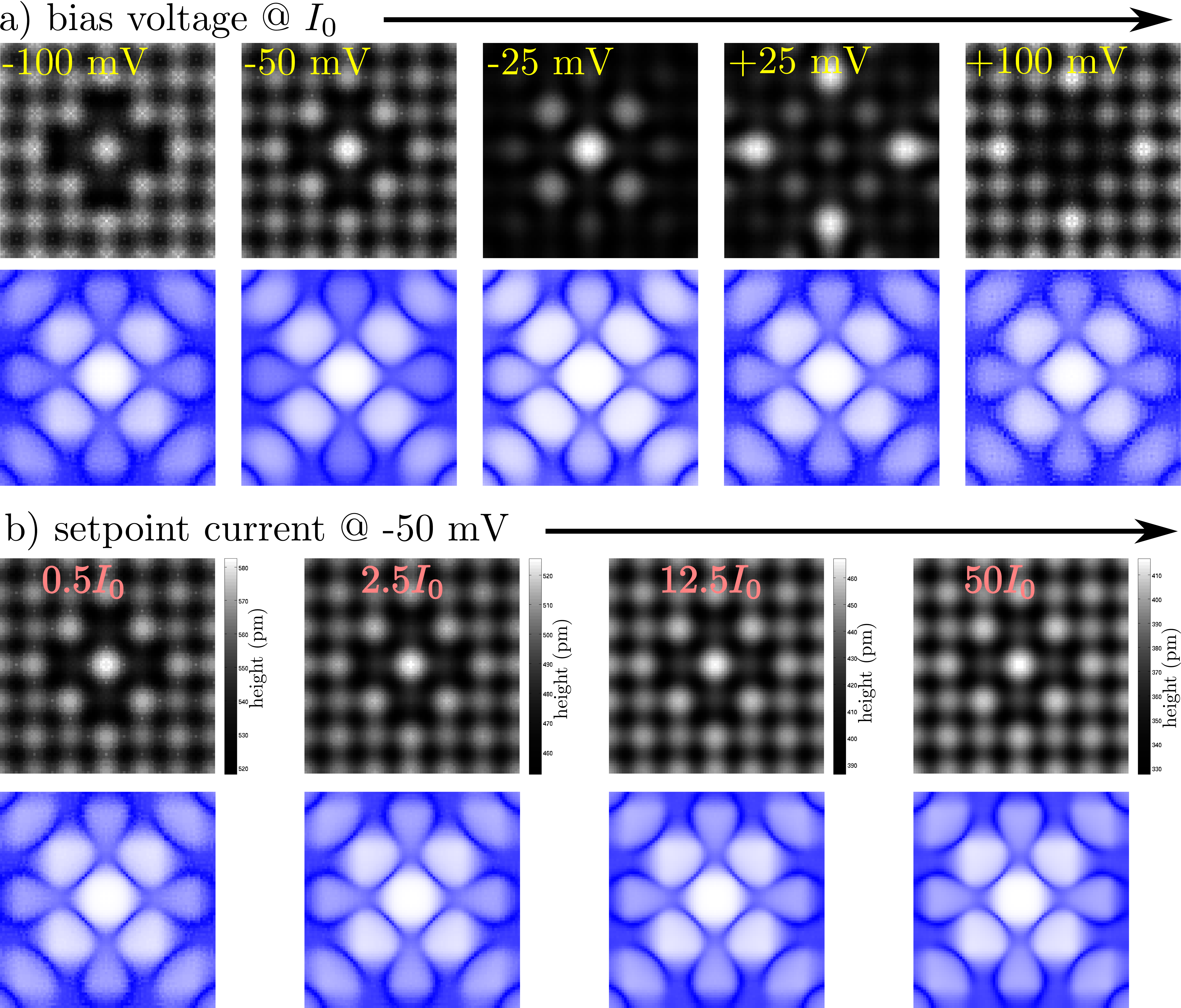}
\caption{Simulation of bias and setpoint current dependence on topographies for NaCCOC (black and white) and topographic conductance maps (blue white, logarithmic scale) at the resonance energy $\Omega=-2.5 \;\text{meV}$:
a) Changing the bias voltage from negative to positive changes the shape of the impurity in the topography (top row), but has little influence on the corresponding topographic conductance map at the resonance energy (bottom row).
b) Changing the setpoint current for the topography does change the overall height, but has almost no influence on the shape as seen in the topography and the topographic conductance maps.}
\label{fig_Zn_CCOC_topo}
\end{figure}

Now, we describe the effects of a strong potential scatterer like Zn in the superconducting state of NaCCOC, and compare it with BSCCO. We use an on-site impurity potential of $U_{0}=-5\;\text{eV}$ to model Zn-like potential scatterer. A very similar value to the impurity potential was obtained from the first principles calculations for Zn substituting a Cu atom in BSCCO \cite{Kreisel15}. The effects of correlations are crudely accounted for by scaling down all hopping parameters by a mass renormalization factor $1/Z = 3 (2.3)$ for BSCCO (NaCCOC) so that the Fermi velocities match with the corresponding experimental values \cite{DamascelliRMP,Zhou03}. The electron filling is set to $n = 0.85$, corresponding to optimal doping. The $d$-wave superconductivity is introduced "by hand" through a pairing interaction  on NN bonds with pair potential $\Gamma = 0.17 (0.15) \,\text{eV}$ for BSCCO (NaCCOC), such that a $d$-wave gap with $\Delta(\mathbf{k}) = \Delta_{0}/2[\cos(k_{x}) - \cos(k_{y})]$, with $\Delta_{0} \approx 33$ meV is obtained for both BSCCO and NaCCOC. Note that the coherence peak positions and therefore the gap maxima are less clear in the tunneling spectra for NaCCOC than for BSCCO; we set the $\Delta_0$'s equal only to highlight the other materials-specific differences between the systems that we study here. Fig. \ref{ldos_strong_impurity}(a) and (d) show the lattice LDOS in the vicinity of a Zn-like impurity in BSCCO and NaCCOC, respectively. A sharp bound state at $\pm \Omega = 4 (3.4) \,\text{meV}$ is observed in the BSCCO (NaCCOC) lattice LDOS spectrum at the NN site, whereas the lattice LDOS at the impurity site is completely suppressed, as one would expect from a strong impurity. As pointed out in Ref. \onlinecite{Kreisel15} and illustrated in Fig. \ref{ldos_strong_impurity}(b) and (c), the experimentally observed \cite{PanZn} spatial distribution of spectrum in Zn-doped BSCCO is recovered in the continuum LDOS, obtained at a height $\approx 5$ {\AA} above the BiO surface using Eq. \ref{eq_cLDOS}, which shows a sharp peak at $\omega = -\Omega$ directly above the impurity site, and large intensity at the central and NNN sites in the spatial map at this resonance energy. More importantly for the focus of this paper, the impurity-induced states in NaCCOC should follow the same pattern, as illustrated in Fig. \ref{ldos_strong_impurity}(e) and (f). This is due to the fact that the Wannier function cut at heights a few Angstroms above the surface (BiO layer in BSCCO and Cl layer in NaCCOC) in both materials are qualitatively very similar, in spite of different surface atoms.  While to our knowledge there are no experiments on Zn impurities in NaCCOC,  the BdG+W calculations predict that Zn in that system should give rise to the same patterns of impurity-induced states as that in  BSCCO.

The qualitative difference between lattice LDOS and continuum LDOS is not only visible in the vicinity of the impurity, but also at sites far from the impurity where a  more $U$-shaped spectrum (see Fig. \ref{ldos_strong_impurity}(b) and (e))  than seen in the usual $V$-shaped lattice LDOS  (see Fig. \ref{ldos_strong_impurity}(a) and (d)) is found.  This unusual $U$-shaped behavior is actually observed in various STM experiments on the overdoped cuprates \cite{Pan2001,Kohsaka08,McElroy05,Alldredge08,Lang2002}. As discussed in detail in the Appendix, this change of low-energy spectrum can be attributed to the particular form of the Wannier function which vanishes directly above the central Cu site ($d$-wave symmetry) and attains its largest value at the NN sites, see Fig. \ref{fig_wannier} (d), (h). A simple calculation (see Appendix) shows that the low-energy continuum LDOS spectrum above a Cu site, at heights a few Angstroms above the surface, varies as $\rho(\mathbf{r},\omega\rightarrow 0) \sim \vert \omega \vert ^3$ yielding a more $U$-shaped structure.   The suppression of the linear-$\omega$ term can be traced back to the occurrence of nonlocal contributions to the continuum Green's function\cite{Choubey2017}.  It is interesting to speculate why STM data seems to indicate a crossover from $U$-shaped (overdoped) to $V$-shaped (underdoped), possibly related to the variation of the range of the Wannier function as correlations grow\cite{Spalek}, but we have not arrived at a completely satisfactory description of this effect.  In any case, the current calculations are applicable only to the more weakly-correlated (overdoped) regime.

So far, we have discussed the calculation of the differential conductance following Eq. (\ref{eq_conductance}) by evaluating the continuum LDOS at fixed height above the surface. Experimentally, the differential conductance maps are usually taken in topographic mode, i.e. while scanning the STM tip over an area by keeping the current $I_0$ at a given bias voltage $V_0$ constant such that the height $z(x,y)$ varies with the position $(x,y)$\cite{PanZn,Hudson2001,Hanaguri2007,Feenstra94}. Theoretically, we can simulate a topograph by solving the equation\cite{Choubey2014}
\begin{equation}
I_0=\frac{4\pi e}{\hbar}\rho_t(0) |M|^2\int_0^{eV_0} d\omega~  \rho({x,y,z(x,y)},\omega)\,,\label{eq_topograph}
\end{equation}
for the height map $z(x,y)$ for given $I_0$ and $V_0$. Indeed, this requires the calculation of the continuum LDOS within a height range and for all energies to carry out the integral. Since we find little difference in the results between the BdG approach and a T-matrix calculation for the relevant quantities (see Appendix), these calculations have been done with the  less computationally expensive T-matrix approach only, as outlined in section \ref{sec_tmatrix}. For the strong potential scatterer $U_0=-5$\,eV, the calculation of topographs has been carried out for NaCCOC only, with the result as presented in Fig. \ref{fig_Zn_CCOC_topo} where in the top rows black and white topographic maps are shown for a fixed current $I_0$ but with varying setpoint voltage $V_0$ (a) and for a fixed setpoint voltage $V_0$ with varying current $I_0$ over several orders of magnitude (b). Indeed, the topographs change with the setpoint bias $V_0$, since the integral in Eq. (\ref{eq_topograph}) is dominated by the real space structure of the impurity resonance at negative energies for $V_0=-25$\,mV and at positive energies for $V_0=+25$\,mV while at large bias voltage (positive or negative) a larger part of the tunneling current $I_0$ comes from electronic states that are dominated by normal state features. The topographs for fixed current bias voltage $V_0$ but varying current $I_0$ (as shown in Fig. \ref{fig_Zn_CCOC_topo}(b)), are very similar except that the overall height is moved down when increasing the current. This is a property of the continuum LDOS and consequently the simulated tunneling conductance being in the exponential limit\cite{Kreisel15}. With the simulated height map $z(x,y)$ in hand, we can also simulate topographic conductance maps $G(x,y,eV)$ as measured in experiment by using $G(x,y,eV)\propto \rho({x,y,z(x,y)},eV)$. For each topography in Fig. \ref{fig_Zn_CCOC_topo}, we also show the topographic conductance maps $G(x,y,\Omega)$ at the resonance energy $\Omega$ with the result that differences are visible in the relative height of the features at the positions above the NN, NNN and NNNN Cu atoms, a consequence of the STM tip being closer to the surface or further away depending on the choice of the setpoint voltage $V_0$.

\subsection{Effects of a magnetic impurity} \label{magImp}

\begin{figure}
\includegraphics[width=1\columnwidth]{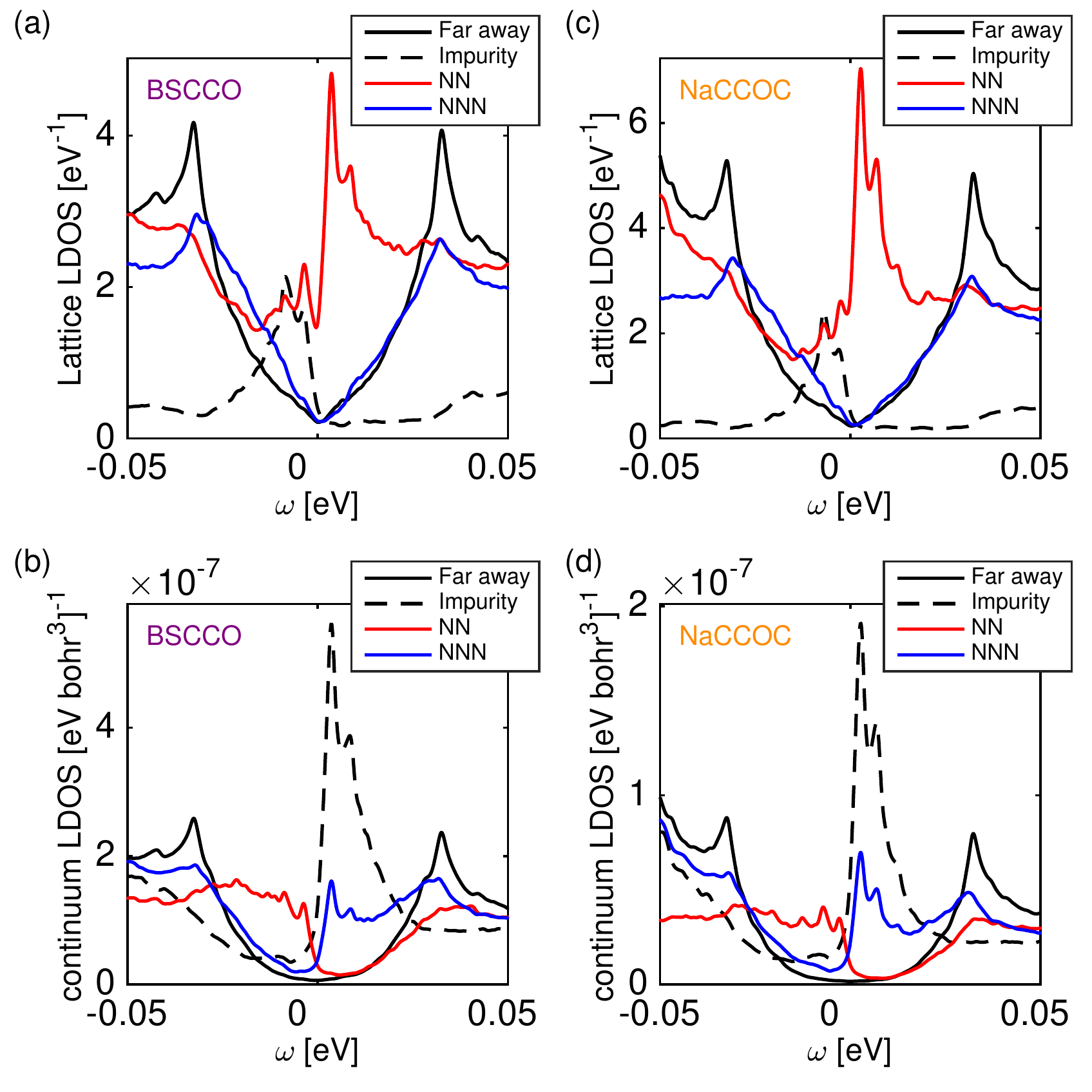}
\caption{(a) Lattice LDOS spectrum around a magnetic impurity, replacing Cu in BSCCO, with on-site potential $U_{0} = 0.6$ eV and on-site exchange coupling $J_{0} = 0.3U_{0}$. Spectrum at a site far from the impurity (black), impurity site (dashed), nearest neighbor site (red), and next-nearest neighbor site (blue) is calculated using $30\times30$ supercells with artificial broadening of 1 meV. (b) Continuum LDOS spectrum at a height approximately 5 {\AA} above the BiO surface. Shown are positions directly above a Cu atom far from impurity (black), at the impurity (dashed), on the nearest neighbor position (red), and on the next-nearest neighbor position (blue). (c),(d) same quantities as in (a) and (b), respectively, for the same magnetic impurity replacing Cu in NaCCOC.}
\label{fig:pointImpuritySpectrum}
\end{figure}

\begin{figure}
\includegraphics[width=1\columnwidth]{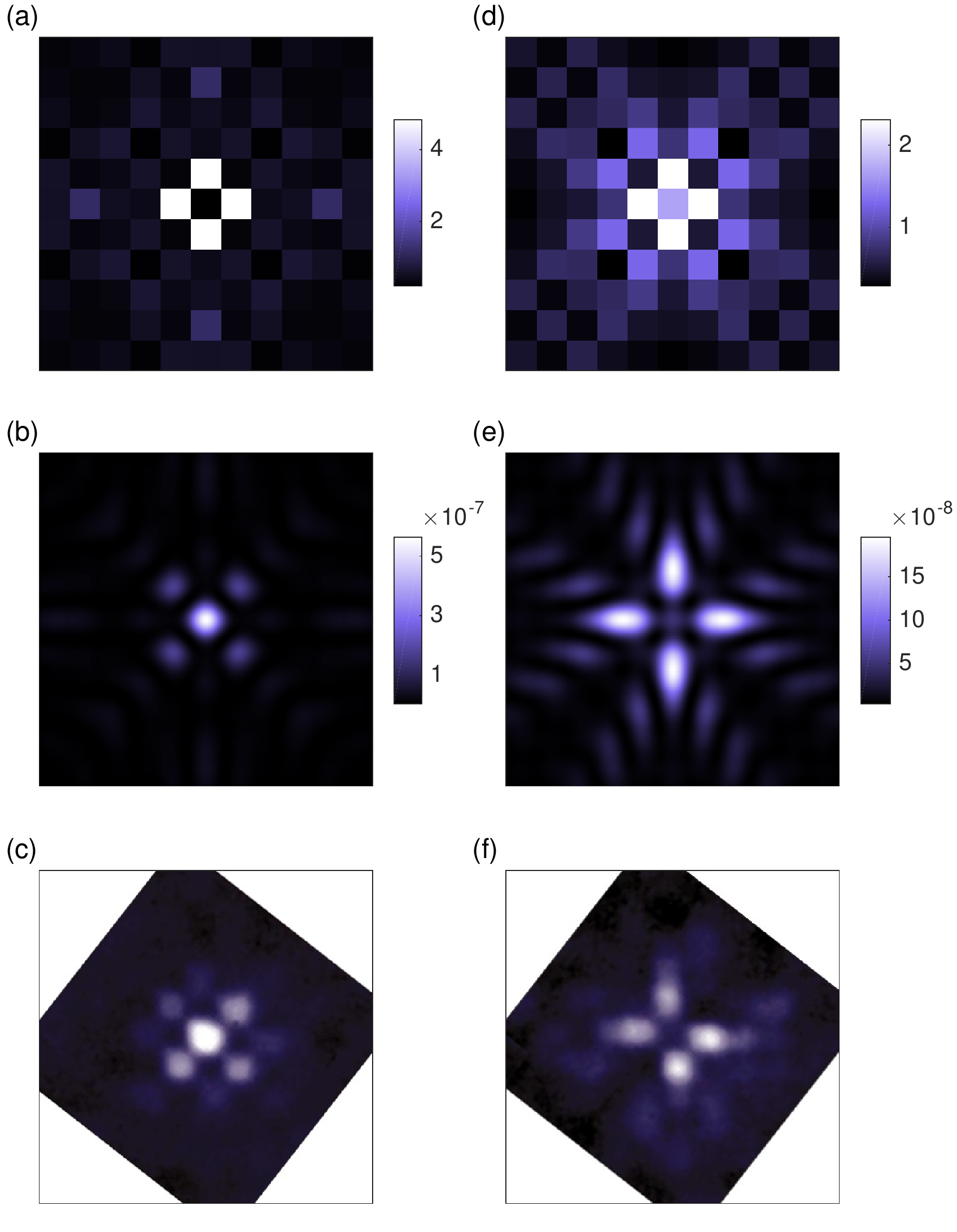}
\caption{(a), (d) Lattice LDOS map around a magnetic impurity in BSCCO in a region comprising $11\times11$ unit cells at the resonant energies $\Omega_{1} = 3.6$ and $-3.6$ meV, respectively. Colorbar values are in the units of eV$^{-1}$. (b), (e) Continuum LDOS maps at $\Omega_{1} = 3.6$ and $-3.6$ meV, respectively, and approximately 5 {\AA} above BiO plane with same area as in (a) and (d). Colorbar values are in the units of eV$^{-1}$bohr$^{-3}$.(c), (f) STM conductance maps at $\Omega_{1}^{\text{expt}} = 9$ meV and $-9$ meV, respectively, reproduced from Ref. \onlinecite{Hudson2001}, rotated to match the orientation in (b) and (c), and cropped to $11\times11$ elementary cells with the Ni impurity located at the center.}
\label{fig:ldosMaps}
\end{figure}

We now study magnetic impurity-induced states induced by Ni in BSCCO and NaCCOC.  Theoretical results on both materials show very similar behavior due to similar tight-binding parameters and Wannier functions. Ni is known to be a weaker impurity than Zn as it does not disrupt superconductivity in its vicinity, and the pure magnetic scattering is thought to be sub-dominant to the pure potential scattering \cite{Hudson2001}. In the absence of magnetic scattering, the resonance peaks are spin degenerate. A small on-site exchange coupling $J_{0}$ lifts this degeneracy as the effective impurity potential experienced by the electrons becomes spin-dependent $V_{\text{imp}}^{\text{eff}} = U_{0} + \sigma J_{0}$, and leads to two spin polarized in-gap resonance states.

Accordingly, in the simplest scenario, we model Ni as a completely local impurity, with on-site potential $U_{0} = 0.6$ eV and exchange coupling $J_{0} = 0.3U_{0}$, placed in the center of a $35\times35$ lattice. Figures \ref{fig:pointImpuritySpectrum}(a) and (c) show the lattice LDOS spectrum at different sites around such an impurity in BSCCO and NaCCOC, respectively. For BSCCO (NaCCOC), impurity induced resonance peaks at $\pm\Omega_{1} = 3.6 (2.7)$ meV and $\pm\Omega_2 = 8.6 (7)$ meV are easily observed in the LDOS spectrum at the impurity site and its NN.  Resonance peaks at $+\Omega_{1}$ and $-\Omega_{2}$ have down-spin polarization whereas those at $+\Omega_{2}$ and $-\Omega_{1}$ have up-spin polarization. By contrast, the spectrum at the NNN site simply follows the bulk LDOS spectrum.  The particular values of $U_{0}$ and $J_{0}$ have been chosen just to clearly demonstrate the spin-splitting of impurity-induced resonance peaks due to on-site magnetic scattering. The experimentally observed sharp resonance peaks close to the gap edge \cite{Hudson2001} can not be reproduced in this simple model, due to significant broadening of peaks close to the gap edge\cite{BalatskyRMP}.
 
In order to directly compare with the STM conductance \cite{Hudson2001}, we must calculate the continuum LDOS at a typical STM tip position. Figures \ref{fig:pointImpuritySpectrum}(b) and (d) show the continuum LDOS spectrum in BSCCO and \mbox{NaCCOC}, respectively, at $z \approx 5$\,\AA{} above the surface at positions which are directly above the impurity, NN site, NNN site, and a distant site. In both materials, similar to the STM result\cite{Hudson2001}, the continuum LDOS exhibits a double peak structure at negative (positive) energies at the positions directly above NN site (impurity and NN sites). Note that this trend is completely missing in the lattice LDOS spectrum (Fig. \ref{fig:pointImpuritySpectrum}(a) and (c)). 

Now we turn to the spatial LDOS maps at the resonance energies in BSCCO. Figure \ref{fig:ldosMaps}(a) shows the lattice LDOS map at $\omega = +\Omega_{1}$ around impurity in a region comprising $11\times 11$ unit cells. The vanishing LDOS at the impurity site and maximum at the NN site observed here is completely opposite to the experimental conductance map at $\omega = +\Omega_{1}^{\text{expt}}$ reproduced in Fig. \ref{fig:ldosMaps}(c). However, the continuum LDOS map $\rho(\mathbf{r},+\Omega_{1})$, at a height $\approx 5$ \AA{} above the BiO plane, shown in Fig. \ref{fig:ldosMaps}(b) compares very well with the experiment. A similar trend holds at the negative energy resonance peak. The continuum LDOS map at $\omega = -\Omega_{1}$ shown in Fig. \ref{fig:ldosMaps}(e) shows excellent agreement with the experimental conductance map at $\omega = -\Omega_{1}^{\text{expt}}$ reproduced in Fig. \ref{fig:ldosMaps}(f). The continuum LDOS maps at $\omega = +\Omega_{2}$ ($-\Omega_{2}$) is found to be very similar to that at $\omega = +\Omega_{1}$ ($-\Omega_{1}$).

Finally, we note that even though the simple on-site magnetic impurity model captures the spin-splitting of impurity resonance peaks and yields correct spatial patterns of LDOS, a careful comparison with the STM conductance spectrum shows that the relative peak heights are reversed. The experimentally observed relative peak heights can not be explained by an on-site magnetic impurity model as the height of the resonance peak decreases and width increases as it moves away from the mid-gap due to increasing hybridization between bulk and impurity states \cite{BalatskyRMP}. In the Appendix, we show that an extended impurity model can yield the experimentally observed relative peak heights, and retain the same continuum LDOS patterns.

\section{Conclusions}

In this work, we have tackled the long-standing question of why STM on cuprate surfaces is apparently universal, despite the fact that chemically quite different barrier layers tend to separate the CuO$_2$ plane from the STM tip above the surface.  Using a recently developed Wannier-based treatment of the local electronic structure combined with $d$-wave superconductivity, we provided a simple explanation of this universality based on the simple form of the Cu $d_{x^2-y^2}$ Wannier function several \AA~above the surface at the tip position.  We showed that while these functions are completely different in the vicinity of the CuO$_2$ plane and inside the barrier layer for two cuprate systems commonly used in STM, BSCCO and NaCCOC, their tails detected by the STM tip are remarkably similar. This fact alone should lead to the nearly identical charge order patterns observed in both materials\cite{Davis_checkerboard2}, provided of course that the charge ordered states in the CuO$_2$ plane are in fact the same.  
   
To illustrate the effect of this ``filtered'' tunneling process and how it affects the spatial conductance patterns observed, we presented again some details of the conductance maps and topographs for a Zn impurity in BSCCO, shown earlier\cite{Kreisel15} to provide a remarkably good agreement with the Zn conductance maps  observed in STM over many years.  As expected from the above discussion, similar calculations for NaCCOC reproduce extremely similar patterns.  This represents a prediction for future experiments, since to our knowledge no experiments on Zn in NaCCOC have been performed.  In addition,  we studied the Ni impurity problem in BSCCO for the first time with the Wannier-based method, obtaining for the BSCCO system conductance maps with extremely good agreement with experimental maps at the resonance energies. Very similar results were obtained for NaCCOC. Some small discrepancies with the height of the two resonances relative to experiment in BSCCO were noted, and  discussed in terms of the range of the Ni impurity potential.  Similar calculations were presented for NaCCOC, again with similar patterns predicted.  

We have furthermore pointed out an interesting problem regarding the low-energy tunneling conductance that has been noted before phenomenologically\cite{Xiang} but not analyzed microscopically.  The $d$-wave form of the Wannier function at the tip position strongly suppresses the linear-$\omega$ bias dependence of the $c$-axis tunneling density of states, leading to a predicted $U$-shaped characteristic.  Indeed, data on overdoped cuprates appears to show this behavior, while for underdoped samples a $V$-shaped spectrum is observed.  We discussed possible effects that might explain the crossover from $U$- to $V$-shaped spectra in terms of enhanced correlations in the underdoped phase.
   
The situation with charge order in the two materials is considerably more subtle, and may depend on a deeper understanding of the origins of charge order in both materials. Recent comparison of inhomogeneous stripe-like charge states of the $t-J$ model, dressed with Wannier functions using the current method\cite{Choubey2017}, suggest that these are indeed of universal origin, and appear identical in the two systems because of the simple form of the Wannier functions above the plane. Future analysis of STM data on  this important problem in cuprate physics should use the Wannier-based method to ensure accurate conclusions regarding correlations in the CuO$_2$ plane itself.

\section{Acknowledgements}

The authors wish to acknowledge useful discussions with J.C. Davis, A. Kostin, W. Ku, and P. Sprau. P.C. and P.J.H. were
supported by Grant No. NSF-DMR-1407502. B.M.A.\ acknowledges support from Lundbeckfond fellowship (grant A9318). Work by T.B. was performed at the Center for nanophase Materials Sciences, a DOE Office of Science user facility. and This research used resources of the National Energy Research Scientific Computing Center, a DOE Office of Science User Facility supported by the Office of Science of the U.S. DOE under Contract No. DE-AC02-05CH11231.

\section*{Appendix}
\subsection*{Homogeneous state continuum LDOS}
Transforming to the momentum-space basis, Eq. (\ref{eq_cLDOSm}) and (\ref{eq_basis}] lead to the following expression for the continuum LDOS in a homogeneous system.
\begin{equation}
\begin{aligned}
\rho_{\sigma}(\mathbf{r},\omega) = \sum_{\mathbf{k}}A_{\sigma}(\mathbf{k},\omega)\vert W_{\mathbf{k}}(\mathbf{r})\vert^{2}.
\end{aligned}
\label{eq:kSpaceLDOS1}
\end{equation}
Here, $W_{\mathbf{k}}(\mathbf{r}) = \sum_{\bf R}w_{\bf R}(\mathbf{r})e^{i\mathbf{k}\cdot\mathbf{R}}$ is the Fourier transform of the Wannier function, and $A_{\sigma}(\mathbf{k},\omega)$ is the spectral function which can be written as. 
\begin{equation}
A_{\sigma}(\mathbf{k},\omega) = \vert u_{\mathbf{k}}\vert^2 \delta(\omega - E_{\mathbf{k}}) + \vert v_{\mathbf{k}}\vert^2 \delta(\omega + E_{\mathbf{k}}).
\label{eq:spectralFunctionK}
\end{equation}
where $E_{\mathbf{k}} = \sqrt[]{\xi_{\mathbf{k}}^2 + \Delta_{\mathbf{k}}^2}$, with  $\xi_{\mathbf{k}}$ being the band energy relative to the Fermi level and superconducting gap $\Delta_{\mathbf{k}} = \Delta_{0}(\cos k_{x} -\cos k_{y})$, and the coherence factors are given as $\vert u_{\mathbf{k}}\vert^2 = \frac{1}{2}\left(1 + \frac{\xi_{\mathbf{k}}}{E_{\mathbf{k}}}\right) = 1 - \vert v_{\mathbf{k}}\vert^2$.  

For the continuum position above the Cu site, $\mathbf{r}_{0} = [0, 0, z]$, at heights $z$ several {\AA} above the BiO plane, the dominant contribution to $W_{\mathbf{k}}(\mathbf{r})$ comes from the the Wannier function at the nearest neighbor sites (see Fig. \ref{fig_wannier}). Accordingly
\begin{equation}
\begin{aligned}
W_{\mathbf{k}}(\mathbf{r}_{0}) \approx w_{0} + 2w_{1}(\cos k_{x} -\cos k_{y})
\end{aligned}
\label{eq:WannierKnn}
\end{equation}
where, $w_{0}$ and $w_{1}$ are the values of the Wannier function at $\mathbf{r} = \mathbf{r}_{0}$, and $\mathbf{r} = \mathbf{r}_{0} + a\hat{x}$, where $a$ is the in-plane lattice constant. The $d_{x^2 - y^2}$-wave symmetry of the Wannier function dictates that $w_{0} = 0$, however, we still keep it to facilitate discussions in following paragraphs. The higher harmonics neglected in Eq. (\ref{eq:WannierKnn}) will not qualitatively change our final conclusions regarding the shape of the continuum LDOS spectrum at low energies.

\begin{figure}[tb]
\begin{center}
\includegraphics[width=\columnwidth]{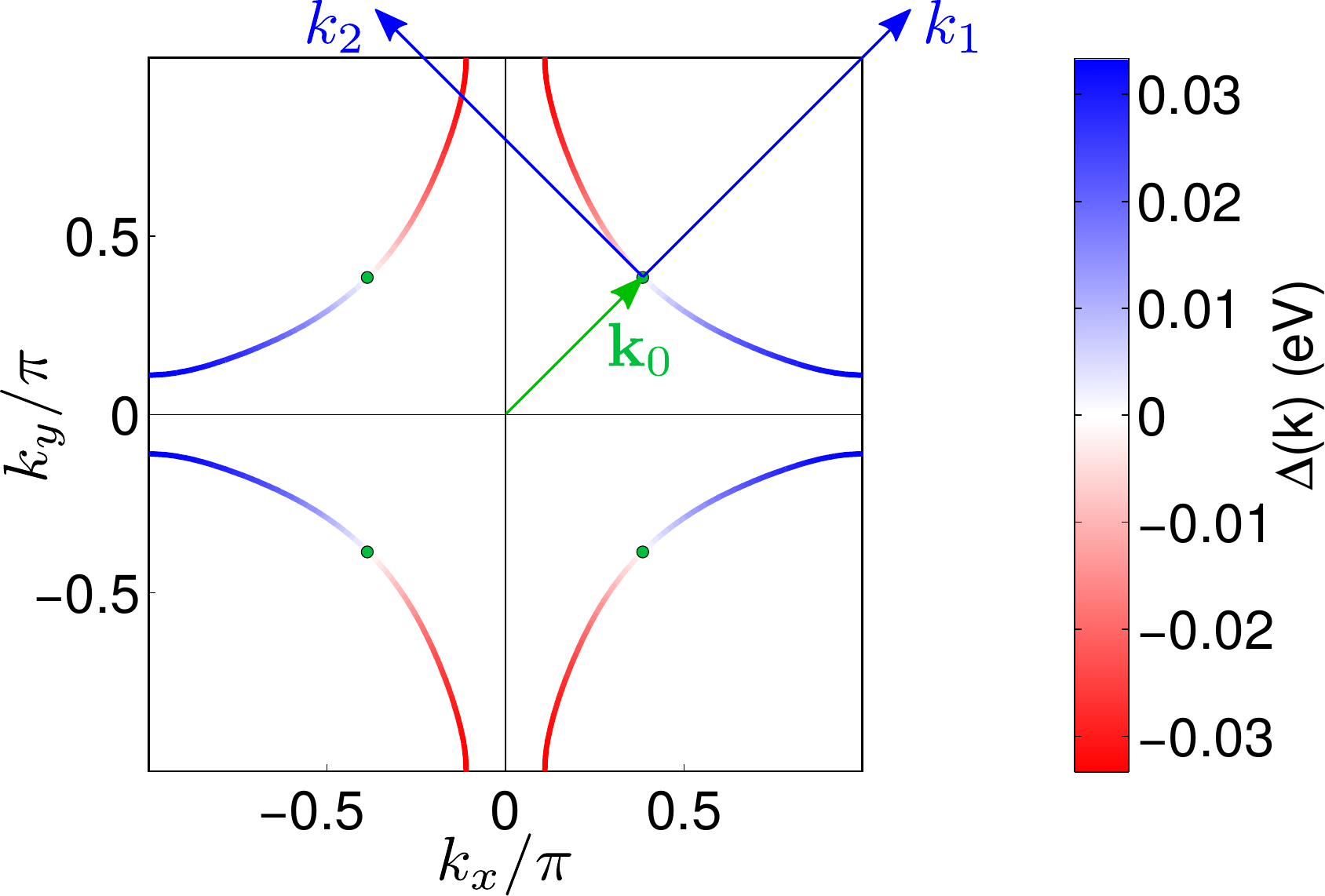}
\caption[Nodal coordinates]
{The d-wave order parameter has $M=4$ nodes (green dots) at the Fermi surface which is shown as an example for BSCCO. The coordinate system $(k_{1},k_{2})$ with axes parallel and perpendicular to a nodal direction (blue arrows) and origin located at the node $\mathbf{k}_{0}$ in the first quadrant (green arrow) is also shown.}
\label{fig:nodalCoordinates}
\end{center}
\end{figure}

Using Eqs. \ref{eq:kSpaceLDOS1}, \ref{eq:spectralFunctionK}, and \ref{eq:WannierKnn}, we can express the continuum LDOS above a Cu site for $\omega > 0$ as
\begin{align}
&\rho_{\sigma}(\mathbf{r}_{0},\omega) = \rho^{00} + \rho^{10} + \rho^{11}\notag\\
&\rho^{00} = \frac{1}{2}w_{0}^2\sum_{\mathbf{k}}\left(1 + \frac{\xi_{\bf{k}}}{E_{\mathbf{k}}}\right)\delta(\omega - E_{\mathbf{k}})\notag\\
&\rho^{10} = 2w_{0}w_{1}\sum_{\mathbf{k}}\left(1 + \frac{\xi_{\mathbf{k}}}{E_{\mathbf{k}}}\right)(\cos k_{x} -\cos k_{y})\delta(\omega - E_{\mathbf{k}})\notag\\
&\rho^{11} = 2w_{1}^2\sum_{\mathbf{k}}\left(1 + \frac{\xi_{\mathbf{k}}}{E_{\mathbf{k}}}\right)(\cos k_{x} -\cos k_{y})^2\delta(\omega - E_{\mathbf{k}}).
\label{eq:ldosCenter}
\end{align}
For $\omega \rightarrow 0$, the dominant contributions to the above integrals will be from the nodal regions of the Brillouin zone. Accordingly, we shift the origin to the the nodal point $\mathbf{k}_{0} = [k_{x}^0, k_{y}^0]$, with $k_{x}^0 = k_{y}^0= \frac{1}{\sqrt[]{2}}{k}_{0}$, and rotate the coordinate axes to $k_{1}$ and $k_{2}$ as shown in Fig. \ref{fig:nodalCoordinates}. Linearizing the band dispersion and gap, we get $\xi_{k} \approx v_{F}k_{2}$ and $\Delta_{\mathbf{k}} \approx v_{\Delta}k_{1}$ where $v_{F}$ is the Fermi velocity at node and $v_{\Delta} = \sqrt[]{2}\Delta_{0}\sin{k_{x}^0}$. With these simplifications, $\rho^{00}$ in Eq. (\ref{eq:ldosCenter}) becomes
\begin{equation}
\begin{aligned}
\rho^{00} = \frac{1}{2}M w_{0}^2 \iint\limits_{\Omega} \frac{d\mathbf{k}}{(2\pi)^2}&\left(1 + \frac{v_{F}k_{2}}{\sqrt[]{v_{\Delta}^2k_{1}^2 + v_{F}^2k_{2}^2}}\right) \\
&\times\delta(\omega - \sqrt[]{v_{\Delta}^2k_{1}^2 + v_{F}^2k_{2}^2}).
\end{aligned}
\end{equation}
where $\Omega$ is a circular region around the nodal point $\mathbf{k}_{0}$ with radius $\Gamma$, and $M = 4$ is the number of nodes. The above integral can be simplified by scaling the coordinates as $k_{1}^{\prime} = v_{\Delta}k_{1}$ and $k_{2}^{\prime} = v_{F}k_{2}$, followed by a transformation to the polar coordinates with $k_{1}^{\prime} = k^{\prime}\cos\theta^{\prime}$ and $k_{2}^{\prime} = k^{\prime}\sin\theta^{\prime}$, resulting into
\begin{align}
\rho^{00} &= \frac{1}{(2\pi)^2}\frac{M w_{0}^2}{2v_{F}v_{\Delta}} \int_{0}^{\Gamma}\int_{0}^{2\pi}k^{\prime}dk^{\prime}d\theta^{\prime}(1 + \sin\theta^{\prime})\delta(\omega - k^{\prime})\notag\\
& = \left(\frac{w_{0}^2 }{\pi v_{F}v_{\Delta}}\right)\omega.
\label{eq:rho00_1}
\end{align}
Proceeding in a similar manner, we can evaluate $\rho^{11}$ as
\begin{align}
\rho^{11} &= 2w_{1}^2M\iint\limits_{\Omega} \frac{d\mathbf{k}}{(2\pi)^2}(1 + \frac{v_{F}k_{2}}{\sqrt[]{v_{\Delta}^2k_{1}^2 + v_{F}^2k_{2}^2}})\notag\\
&  \qquad \qquad \quad \times(2\,\,\sqrt[]{2}t\,k_1\, \sin{k_{x}^0})^{2}\delta(\omega - \sqrt[]{v_{\Delta}^2k_{1}^2 + v_{F}^2k_{2}^2})\notag\\
&=\frac{4M w_{1}^2\sin^{2}k_{x}^0 }{(2\pi)^2v_{F}v_{\Delta}^{3}}\int_{0}^{\Gamma}\int_{0}^{2\pi}k^{\prime}dk^{\prime}d\theta^{\prime}\delta(\omega - k^{\prime})\notag\\
& \qquad \qquad \qquad \qquad \qquad  \times\left(k^{\prime 2}\cos^2\theta^{\prime} + \sin\theta^{\prime}\cos^2\theta^{\prime}\right)\notag\\
&=\left(\frac{4 w_{1}^2\sin^{2}k_{x}^0 }{\pi^3v_{F}v_{\Delta}^{3}}\right)\omega^{3}.
\end{align}
Finally, $\rho^{10}$ turns out to be zero due to vanishing angular integrals. Thus, 
\begin{equation}
\begin{aligned}
&\rho_{\sigma}(\mathbf{r}_{0},\omega) = a_{0}\omega + a_{1}\omega^{3},\\
&a_{0} = \frac{w_{0}^2 }{\pi v_{F}v_{\Delta}}, \quad a_{1} = \frac{4 w_{1}^2\sin^{2}k_{x}^0 }{\pi^3v_{F}v_{\Delta}^{3}}.
\end{aligned}
\label{eq:rho}
\end{equation}
The coefficient of the linear in $\omega$ term vanishes as $w_{0} = 0$ due to $d_{x^2 - y^2}$ symmetry of the Wannier function, however, the coefficient of $\omega^3$ term is non-zero, and yields a $U$-shaped spectrum at low energies.  The equivalent result was obtained earlier using a purely phenomenological model for the hopping between cuprate layers\cite{Xiang}. Setting $W_{\mathbf{k}}(\mathbf{r}) = 1$ in Eq. (\ref{eq:kSpaceLDOS1}) [or equivalently $w_{0} = 1$, $w_{1} = 0$ in Eq. (\ref{eq:WannierKnn})] yields the lattice LDOS $N(\omega)$ which, from Eq. (\ref{eq:rho}), turns out to vary linearly with $\omega$, leading to the well-known $V$-shaped spectrum at low energies. 
\subsection*{{Details of continuum LDOS spectrum in the vicinity of an impurity}}
\begin{figure}[tb]
\includegraphics[width=1\columnwidth]{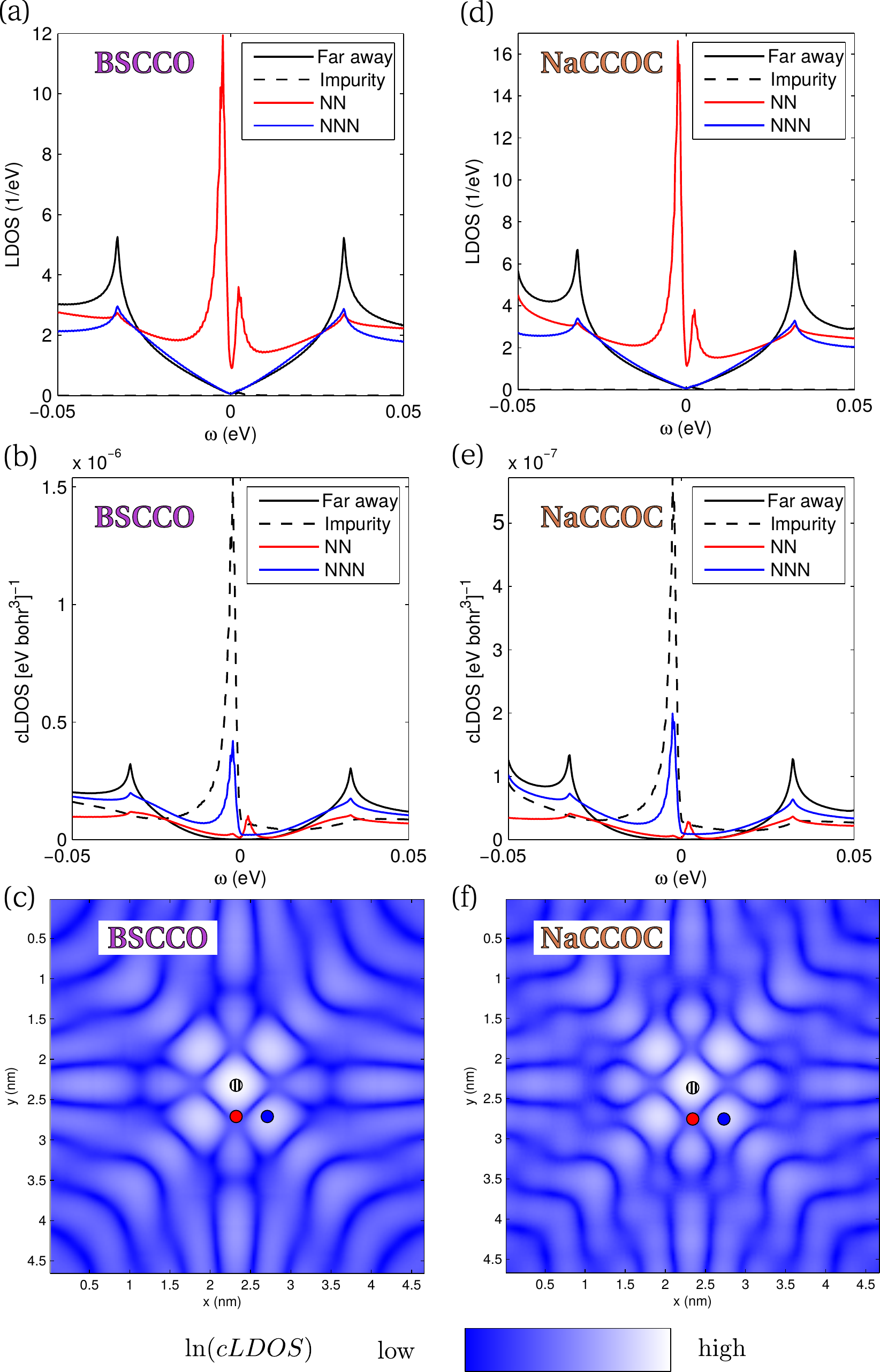}
 \caption{Results for a strong potential scatterer $U_0=-5$\,eV using the T-matrix formalism with $2500 \times 2500$ k-points and a smearing of $\eta=0.25$\,meV. As in Fig. \ref{ldos_strong_impurity}, the lattice LDOS, the continuum LDOS (cLDOS) spectrum, and cLDOS map at a height $\approx 5$ \AA{} above the surface for the resonance energy is plotted both for BSCCO and NaCCOC.}
 \label{fig_compare_T_BdG}
\end{figure}

\begin{figure}[tb]
 \includegraphics[width=1\columnwidth]{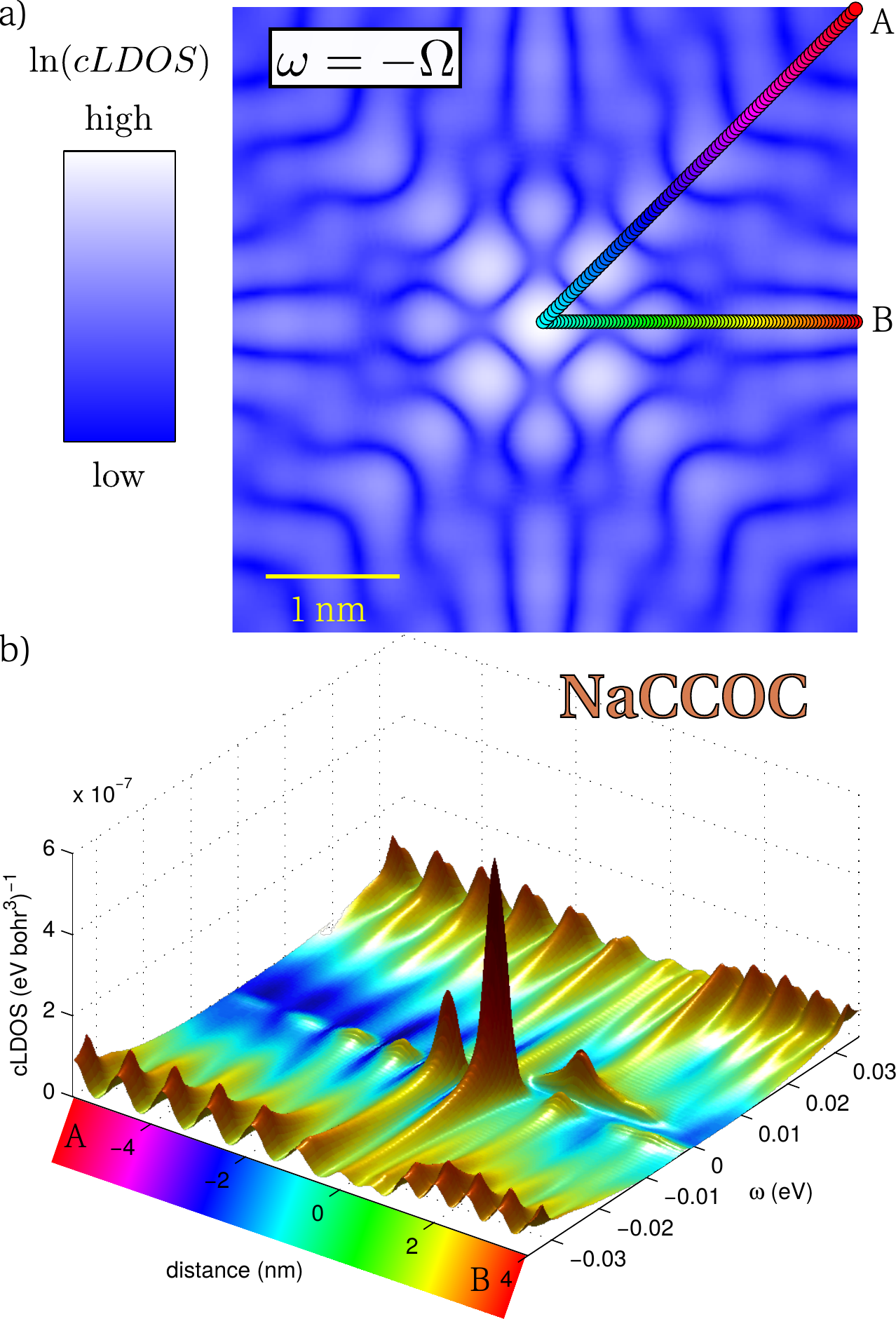}
\caption{Spatial evolution of the spectra close to a strong impurity in NaCCOC: a) continuum LDOS (cLDOS) map from a T-matrix calculation, plotted on an area of 12$\times$12 elementary cells, at the resonance energy $\omega=-\Omega$. Spectra are then  plotted  in b) along the path shown in a), from point A to the impurity position and then along the Cu-Cu bond direction to point B.  The strong impurity resonance peak at the impurity position is visible, along with  further peaks at $\omega=-\Omega$ along the diagonal and oscillating peaks at $\omega=+\Omega$ and $\omega=-\Omega$ along the Cu-Cu bond direction.}
 \label{fig_cut_CCOC}
\end{figure}

As argued already in the main text, the results for various quantities in presence of an impurity do not significantly depend on whether the T-matrix formalism or the self-consistent BdG method is used.
The latter is indeed numerically more demanding because self-consistence in the densities and the superconducting order parameter is required and diagonalization of large matrices is needed.
To show explicitly the correspondence of the results, we present in Fig. \ref{fig_compare_T_BdG} the same quantities as plotted in Fig. \ref{ldos_strong_impurity}, but calculated within the T-matrix formalism.
Keeping this in mind, we can safely look at T-matrix results only for the continuum LDOS spectra in the vicinity of a strong impurity.
The different shape of the impurity pattern in the continuum LDOS at resonance  is also reflected in the spectra taken at different points in real space. For example, if one plots the spectrum  along a path away from the impurity, one sees that along the diagonals of the continuum LDOS maps at positive resonance energy $\Omega$, no intensity can be seen, while  along the Cu-Cu bond direction, the conductance shows a periodic variation. The situation is different for the conductance at negative resonance energy $-\Omega$, where strong variations are seen at the diagonals, and weak variations are seen along the Cu-Cu bonds direction. The same behavior is also demonstrated in Fig. \ref{fig_cut_CCOC} where such a path along the diagonal to the impurity and then along the Cu-Cu bond direction away from the impurity is defined in (a). The spectra along this path as shown in (b) then exhibit periodic modulations of the $-\Omega$ peak along the diagonal and alternating (decreasing) peaks at $+\Omega$ and $-\Omega$ when moving away from the impurity along the Cu-Cu bond direction towards point B.

\subsection*{Normalized conductance maps}
\begin{figure}[tb]
 \includegraphics[width=1\columnwidth]{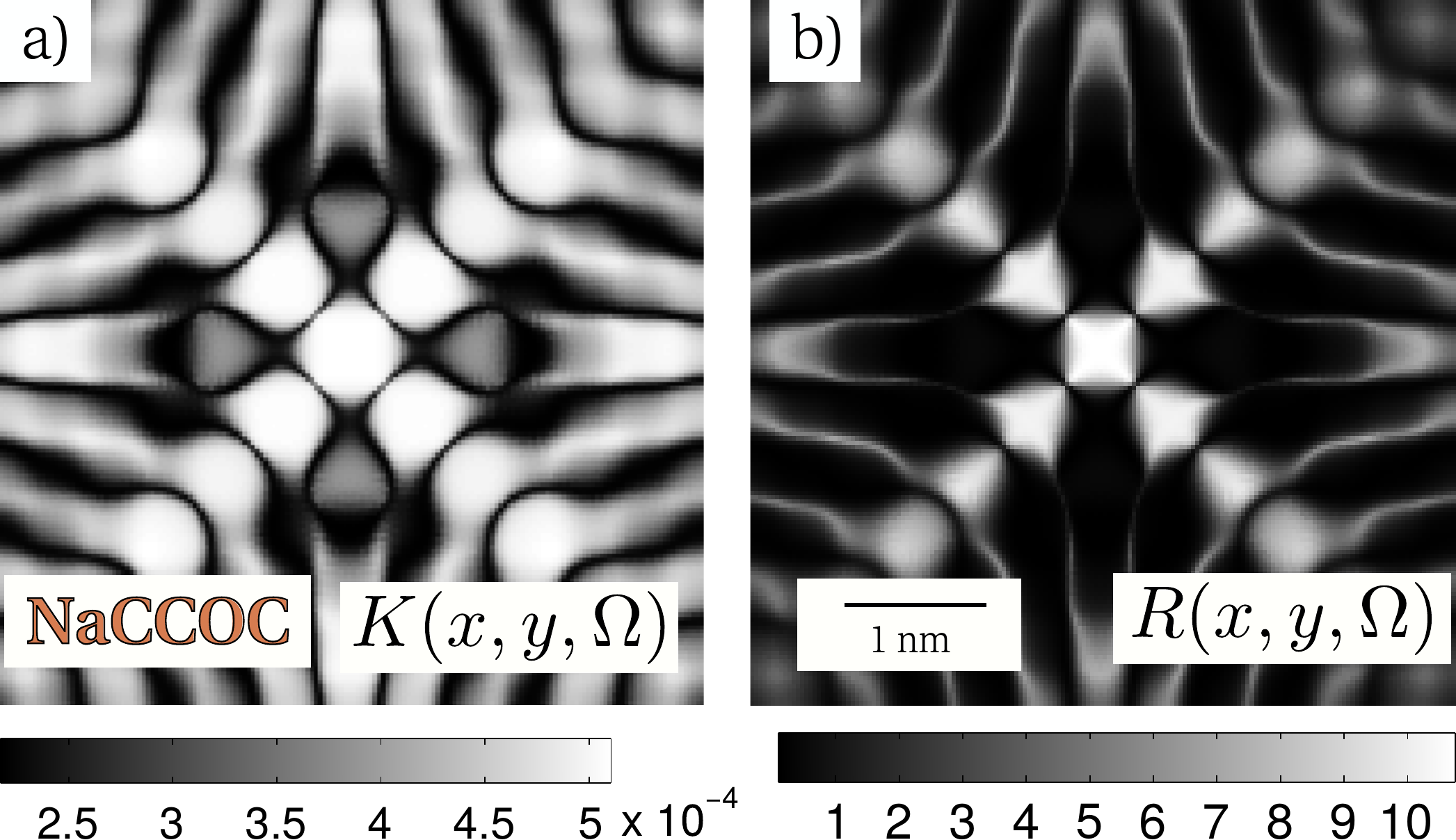}
 \caption{(a) Calculated K-map of a strong impurity and (b) R-map of the same impurity in NaCCOC. The energy is fixed to the resonance energy $\Omega$ and the size of the field of view is 12 unit cells.}
 \label{fig_KR}
\end{figure}
In the main text, we discussed that differential conductance maps slightly depend on the experimental conditions at which the tunneling junction is set up to scan along the surface.
A possible way to eliminate this dependence on the experimental parameters ($I_0$ and $V_0$) is to consider the normalized conductance\cite{Feenstra94}
\begin{equation}
K(x,y,eV)=V \frac{ \frac{dI}{dV}(x,y,eV)}{I(x,y,eV)},
\end{equation}
which is the ratio of the measured differential conductance and the (at the same time) registered current $I(x,y,eV)$, normalized by the voltage $V$. As pointed out in Ref. \onlinecite{Feenstra94}, this quantity is also independent of whether the data has been taken at constant height conditions or constant current conditions (topographic mode) and should be examined in linear scale instead of logarithmic as the conductance maps.

Following our approach for the calculation of differential conductance and current, Eqs. (\ref{eq_conductance}) and (\ref{eq_topograph}), we can theoretically obtain this quantity
\begin{align}
 K(x,y,eV)&=V \frac{\rho({x,y,z(x,y),eV})}{\int_0^{eV} d\omega  \rho({x,y,z(x,y)},\omega)}\notag\\
 &=V \frac{\rho_p({x,y,eV})}{\int_0^{eV} d\omega  \rho_p({x,y},\omega)}.\label{Eq:K-map2}
\end{align}
In the exponential limit, the continuum LDOS will have an exponential dependence on the height $\rho({x,y,z(x,y)})\approx e^{-\alpha z(x,y)} \rho_p({x,y})$ that dropped out from the formula (\ref{Eq:K-map2}) with the last equal sign, such that it is not necessary to calculate the topographic map $z(x,y)$. For quantities that are particle-hole symmetric, a similar removal of the setpoint effect can be achieved in to ways: A division of the conductance by the conductance at negative bias yield the quantity
\begin{equation}
 Z(x,y,eV)=\frac{ \frac{dI}{dV}(x,y,eV)}{\frac{dI}{dV}(x,y,-eV)}
\end{equation}
Finally, one can also divide the current at positive bias by the current at negative bias to get
\begin{equation}
 R(x,y,eV)=\frac{I(x,y,eV)}{I(x,y,-eV)},
\end{equation}
which shows similar features. An example of such a R-map around a strong $U_0=-5$\;eV impurity is shown in Fig. \ref{fig_KR}. The $Z$-map is extremely similar.

\subsection*{Extended magnetic impurity model}

\begin{figure}[t]
\includegraphics[width=1\columnwidth]{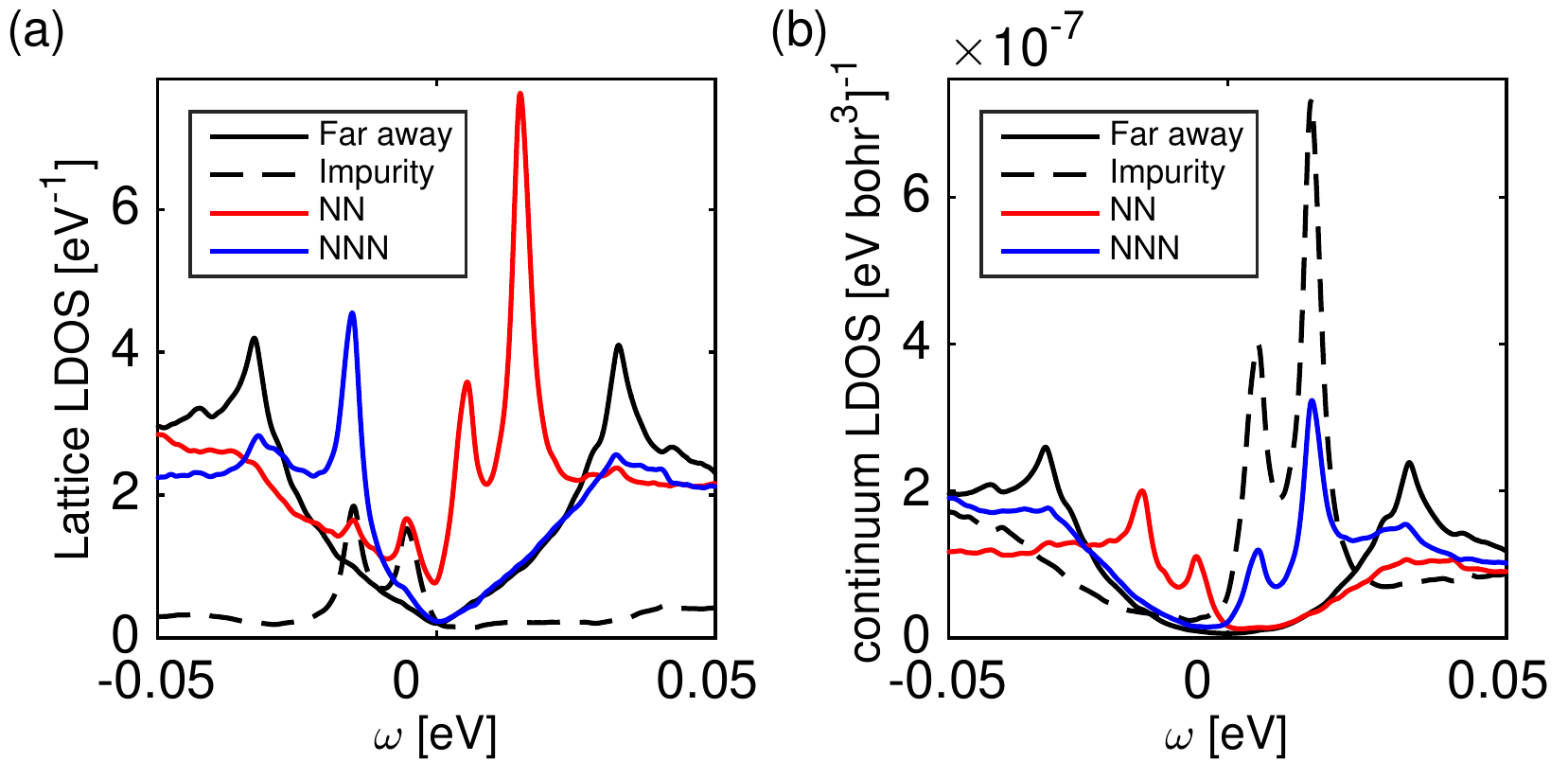}
\caption{(a) Lattice LDOS spectrum around a magnetic impurity, replacing Cu in BSCCO, with on-site potential $U_{0} = 0.6$ eV, NN potential $U_{NN} = 0.5U_{0}$ and NN exchange coupling $J_{NN} = 0.5U_{0}$. Spectrum at a site far from the impurity (black), impurity site (dashed), nearest neighbor site (red), and next-nearest neighbor site (blue) is calculated using $30\times30$ supercells with artificial broadening of 1 meV. (b) Continuum LDOS spectrum at a height $\approx 5$ {\AA} above the BiO surface. Shown are positions directly above a Cu atom far from impurity (black), at the impurity (dashed), on the nearest neighbor position (red), and on the next-nearest neighbor position (blue).}
\label{fig:extendedImpuritySpectrum}
\end{figure}
Although the splitting of peaks due to magnetic scattering is captured in the simple model of a point-like impurity as discussed in section \ref{magImp}, a careful comparison with the experimental conductance spectrum shows that the relative heights of the peaks are reversed. In the simple on-site impurity model, the height of the resonance peak decreases and width increases as it moves away from the mid-gap due to increasing hybridization between bulk and impurity states \cite{BalatskyRMP}. Thus the experimentally observed relative peak heights can not be explained by such model. To this end, we find that an extended impurity model with on-site potential $U_{0} = 0.6$ eV, NN potential $U_{NN} = 0.3$ eV, and NN exchange coupling $J_{NN} = U_{NN}$ can lead to the desired trend\cite{Tang2002,Vojta}. This impurity model yields sharp resonance peaks at $\pm\Omega_{1} = 4.2$ meV and $\pm\Omega_{2} = 18.6$ meV in the lattice and continuum LDOS spectrum as shown in Figure \ref{fig:extendedImpuritySpectrum}(a) and (b), respectively. Clearly, resonance peaks at $\pm\Omega_{2}$ are higher than those at $\pm\Omega_{1}$. Moreover, similar to the experiment and also to the case of on-site impurity model, continuum LDOS shown in Figure \ref{fig:extendedImpuritySpectrum}(b) displays switching of resonance peaks from positive to negative biases and then back to positive biases as one moves from impurity to NN and then to NNN sites. Thus, the extended impurity model captures most of the features of STM results \cite{Hudson2001}; however, the microscopic origin of such a model still needs to be investigated.

\subsection*{Details of first principles calculations}

The DFT calculations were performed within the generalized gradient approximation (GGA) using the Perdew-Burke-Ernzerhof exchange correlation scheme5 and projector augmented wave potentials\cite{Bloechl} as implemented in VASP\cite{Kresse,KresseJoubert}. We used a 7x7x1 k-point grid and a relatively high kinetic energy cutoff of 2000 eV to ensure high quality Wannier functions. To capture the Wannier functions we projected the Cu-$d_{x^2-y^2}$ orbital on the bands within [-3,3]eV. Furthermore we used Wannier90\cite{Mostofi2014} in which we set num\_iter=0 to retain the correct symmetry of the $d_{x^2-y^2}$ orbital and dis\_num\_iter=1000. For the \CCOC Wannier calculation we set dis\_froz\_min =-0.9 and dis\_froz\_max=2.5 to better capture the bands near the Fermi energy. To reduce the two band Wannier function based Hamiltonian of \BSCCO to a single band Hamiltonian we simply cut the (relatively weak) hopping elements that couple two Cu planes. Atomic and 3D Wannier function images were produced with the VESTA program\cite{MommaIzumi}.

\bibliography{literature_STM}{}

\end{document}